\documentclass[final]{cas-dc}

\makeatletter
\def\ps@headings{
  \let\@oddfoot\@empty
  \let\@evenfoot\@empty
}
\def\ps@myplain{
  \let\@oddfoot\@empty
  \let\@evenfoot\@empty
}
\makeatother

\makeatletter
\def\@elsevier{}
\def\@journal{}
\def\ps@first{
  \let\@oddfoot\@empty
  \let\@evenfoot\@empty
}
\makeatother

\usepackage[numbers]{natbib}

\usepackage{siunitx}
\usepackage{threeparttable}

\usepackage[mathlines, switch]{lineno}

\def\tsc#1{\csdef{#1}{\textsc{\lowercase{#1}}\xspace}}
\tsc{WGM}
\tsc{QE}

\begin{document}
\let\WriteBookmarks\relax
\def\floatpagepagefraction{1}
\def\textpagefraction{.001}

\shorttitle{}    

\shortauthors{}  

\title [mode = title]{Fabrication and characterization of boron-terminated tetravacancies in monolayer hBN using STEM, EELS and electron ptychography}

\author[1,2,3]{Dana O. Byrne}
\cormark[1]

\affiliation[1]{organization={Department of Chemistry},
            addressline={University of California}, 
            city={Berkeley},
            state={CA},
            postcode={94720}, 
            country={USA}}

            \affiliation[2]{organization={Department of Materials Science and Engineering},
            addressline={University of California}, 
            city={Berkeley},
            state={CA},
            postcode={94720}, 
            country={USA}}

             \affiliation[3]{organization={Molecular Foundry},
            addressline={Lawrence Berkeley National Laboratory}, 
            city={Berkeley},
            state={CA},
            postcode={94720}, 
            country={USA}}

            \ead{dana_byrne@berkeley.edu}

\author[3]{Stephanie M. Ribet}

\author[4,5]{Demie Kepaptsoglou}

\affiliation[4]{organization={SuperSTEM Laboratory},
            addressline={SciTech Daresbury Campus}, 
            city={Daresbury},
            postcode={WA4 4AD}, 
            country={UK}}

            \affiliation[5]{organization={School of Physics, Engineering and Technology \& JEOL Nanocentre},
            addressline={University of York}, 
            city={Heslington},
            postcode={YO10 5DD}, 
            country={UK}}
            
\author[4,6]{Quentin M. Ramasse}

            \affiliation[6]{organization={School of Chemical and Process Engineering \& School of Physics and Astronomy},
            addressline={University of Leeds}, 
            postcode={LS2 9JT}, 
            country={UK}}

\author[7]{Colin Ophus}

\affiliation[7]{organization={Department of Materials Science and Engineering},
            addressline={Stanford University},
            state={CA},
            postcode={94305}, 
            country={USA}}
            
\author[2,3,8]{Frances I. Allen}
\cormark[1]

             \affiliation[8]{organization={California Institute for Quantitative Biosciences},
            addressline={University of California}, 
            city={Berkeley},
            state={CA},
            postcode={94720}, 
            country={USA}}

\ead{francesallen@berkeley.edu}

\cortext[1]{Corresponding authors}

\begin{abstract}
Tetravacancies in monolayer hexagonal boron nitride (hBN) with consistent edge termination (boron or nitrogen) form triangular nanopores with electrostatic potentials that can be leveraged for applications such as selective ion transport and neuromorphic computing. In order to quantitatively predict the properties of these structures, an atomic-level understanding of their local electronic and chemical environments is required. Moreover, robust methods for their precision manufacture are needed. Here were use electron irradiation in a scanning transmission electron microscope (STEM) at high dose rate to drive the formation of boron-terminated tetravacancies in monolayer hBN. Characterization of the defects is achieved using aberration-corrected STEM, monochromated electron energy-loss spectroscopy (EELS), and electron ptychography. Z-contrast in STEM and chemical fingerprinting by core-loss EELS enable identification of the edge terminations, while electron ptychography gives insight into structural relaxation of the tetravacancies and provides evidence of enhanced electron density around the defect perimeters indicative of bonding effects. 
\end{abstract}


\maketitle


\section{Introduction}\label{}

With the advent of low-voltage aberration-corrected scanning/transmission electron microscopy (S/TEM), atomic-resolution imaging of the delicate monolayers of 2D materials has become possible~\citep{Meyer2008,Krivanek2010}. However, it was soon discovered that significant re-arrangement and removal of atoms from the lattice can still occur~\citep{Girit2009}, even at beam energies well below elastic knock-on thresholds, due to inelastic electron scattering effects such as radiolysis~\citep{Susi2019}. Although sample damage during imaging typically needs to be minimized, there is also interest in leveraging this damage in a controlled way for defect engineering on the single atom level~\citep{Susi2017,Hudak2018,Su2019,Susi2022,Boebinger2023}. 

The first high-resolution studies of electron-irradiation-induced atomic rearrangements in a 2D material focused on graphene, reporting the formation of holes with the so-called zigzag and armchair edges~\citep{Girit2009}. Soon after, experiments with 2D hexagonal boron nitride (hBN) were conducted, demonstrating the formation of perfect triangular-shaped multi-atom vacancy defects under the electron beam~\citep{Jin2009,Meyer2009,Alem2009}. The regular geometric shape is due to preferential ejection of one atomic species over the other, which drives the expanding defect to follow the triangular sublattice of either element. 

While initial reports of triangular vacancy defect growth in 2D hBN determined preferential boron atom ejection (creating nitrogen-terminated defects)~\cite{Jin2009,Alem2009,Ryu2015}, other studies have reported preferential nitrogen atom ejection~\cite{Cretu2015a}, or even, no particular atomic preference at all~\cite{Cretu2015,Pham2016,Gilbert2017,Javed2024}. The key parameters affecting the energetics of the defect expansion process are beam energy, sample temperature, electron dose rate, and the presence (or lack) of residual species in the vacuum. Clearly, these factors need to be thoroughly understood in order for precision engineering of defects in the electron microscope to become a predictable and reliable enterprise.

Tetravacancies in 2D hBN with consistent edge termination (lined with either boron or nitrogen atoms) are of particular interest in membrane applications that leverage the electrostatic potential inside such pores. These applications include selective filtration via mechanosensitive ion transport~\cite{Noh2024} and neuromorphic computing via memristive response~\cite{Noh2024a}. In order to quantitatively predict the behavior of the nanopores, their atomic, electronic and chemical structure need to be precisely known. Here, advanced electron microscopy can lead the way, using aberration-corrected STEM at low-kV for defect identification at the single atom level~\cite{Guo2014}, high energy-resolution electron energy-loss spectroscopy (EELS) to probe the electronic and chemical structure of the defects~\cite{Zhou2012,Ramasse2013,Hage2019,Hage2020}, and electron ptychography for dose-efficient mapping of atom positions and properties with deep sub-\text{\AA ngstr\"om} resolution by quantitative phase retrieval of the electron exit wave~\cite{Jiang2018,Ophus2019}. 

In this work, we employ aberration-corrected STEM, monochromated EELS and electron ptychography to fabricate and analyze vacancy defect structures in hBN monolayers. All experiments are performed in UHV at \qty{60}{kV} with the samples held at room temperature. When scanning at high dose rates, we find that boron-terminated tetravacancies are preferentially formed. The defect edge terminations are first determined based on Z-contrast in high angle annular dark-field (HAADF) STEM images, and also using chemical fingerprinting via monochromated STEM-EELS of the boron K-edge. Atom maps obtained by electron ptychography reveal distortions around the defect edges indicative of structural stabilization, although artifacts from sample charging during the scans must also be considered. The ptychographic reconstructions also show enhanced phase shifts around the defect perimeters, the potential origins of which are discussed. 

\section{Experimental}\label{}

\subsection{Preparation of hBN monolayers}

Monolayer hBN, grown by chemical vapor deposition on copper (Grolltex Inc.), was transferred onto custom holey silicon nitride (SiN$_x$) TEM grids using a wet transfer electrochemical delamination method~\cite{Byrne2025}. For this, small sections of the hBN on copper were cut out and spin-coated with 11\% polymethyl methacrylate (PMMA) in anisole. The copper/hBN/PMMA films were then clipped to the anode of a two electrode system with NaCl/water electrolyte and a glassy carbon cathode. A \qty{5}{V} bias was applied to generate hydrogen bubbles at the anode, which gently removes the hBN/PMMA layer from the copper. The hBN/PMMA film was washed several times in deionized water and transferred onto the custom SiN$_x$ TEM grids. Finally, the PMMA was removed with an acetone drip. 

In the experiments described here, we worked with both ``pristine'' regions of the hBN (native defects only) and regions that had been irradiated with helium ions to create higher densities of single atom vacancy defect seeds for subsequent expansion upon electron irradiation in STEM. The ion irradiation was performed site selectively using a Zeiss ORION NanoFab multibeam He-Ne-Ga focused ion beam microscope, using the helium ion beam in a low-dose raster~\cite{Byrne2025,Byrne2024}. A typical irradiation dose was \qty{10}{ions/nm\textsuperscript{2}} and the helium ion beam energy was set to \qty{25}{keV}. The ion-irradiated samples were also air-annealed to remove additional hydrocarbon surface contamination that formed on those specimens and to promote the growth of larger defect structures, as described in our previous study~\cite{Byrne2025}. 

\subsection{HAADF-STEM and monochromated STEM-EELS}

Electron microscopy was performed using a Nion UltraSTEM 100MC ``Hermes'' aberration-corrected microscope operated at \qty{60}{kV} using a convergence semi-angle of \qty{31}{mrad}, corresponding to $\sim$\qty{1}{\angstrom} electron probe diameter. The vacuum at the sample ranged from $\sim$\qtyrange{1.2e-9}{7.8e-10}{Torr}, depending on the time that had elapsed after sample insertion (this variation is unlikely to result in significant differences in the partial pressures of molecules contributing to chemical etching). All experiments were performed at room temperature. Defect growth was observed both during lower dose rate survey imaging and in the higher dose rate EELS (and later ptychography) acquisitions. 

For the lower magnification HAADF-STEM survey imaging, typical parameters were a beam current of \qty{6}{pA}, pixel dwell time of \qty{240}{\micro s}, and scan spacing of \qty{0.02}{nm}. A typical scan size was 512$\times$512 pixels (field of view \qty{10.24}{nm}), giving a time-averaged dose rate of \qty{3.6e5}{e/nm^2/s} (calculated from the beam current divided by the scan area). The corresponding accumulated dose was \qty{2.3e7}{e/nm^2}. 

Most high magnification HAADF-STEM scans were acquired using a beam current of \qty{6}{pA}, pixel dwell time of \qty{4.9}{ms}, and scan spacing of \qty{0.03}{nm}, for a series of five or more frames. The typical scan size was 50$\times$50 pixels (field of view \qty{1.5}{nm}), giving a time-averaged dose rate of \qty{1.7e7}{e/nm^2/s}. The corresponding total dose for a single frame was \qty{2.0e8}{e/nm^2}. Image series with minimal defect movement between frames were identified, manually drift corrected using DigitalMicrograph\textregistered\ software, and finally averaged using custom Python code to increase the signal-to-noise ratio in the final image. 

In parallel with the high magnification HAADF-STEM acquisitions, an EEL spectrum was collected for every pixel in the scan using the Nion IRIS spectrometer equipped with a Dectris ELA hybrid-pixel detector. The beam current used for the EELS results shown in this paper was \qty{30}{pA}, corresponding to a time-averaged dose rate of \qty{1.1e8}{e/nm^2/s} for a 41$\times$45 pixel scan. 
A collection angle of \qty{44}{mrad} was used and the energy dispersion was \qty{0.1}{eV/channel}. Here, we focus primarily on the boron K-edge. 

The STEM-EELS data were analyzed using the open-source \texttt{HyperSpy} package~\cite{HyperSpy} and custom Python code. This involved spatial drift correction of the frames in the spectrum image using the accompanying HAADF series, image and spectrum averaging, spatial binning, spectrum smoothing with a total variation function, and then non-negative matrix factorization (NNMF) using the \texttt{Scipy} package~\cite{2020SciPy-NMeth} to identify and map spectral features in the vicinity of the onset of the boron K-edge. NNMF score images were used to generate boolean masks, which allowed for spatial separation of features for spectral plotting (see Fig.~S1). The final spectra presented here have been normalized to the $\sigma$* peak of the boron K-edge (full spectral range shown in Fig.~S2). All spectra were shifted in the energy axis to align with the energy-loss value of the B--N $\pi$* peak obtained in~\cite{Byrne2024}. Shifts in the energy axis did occur over the span of several hours, so the calibrations are never absolute, but internally all efforts are made to ensure spectra are comparable across datasets. 

\subsection{Electron ptychography}

Data collection for the ptychography was also performed using the Nion UltraSTEM microscope operated at \qty{60}{kV} with a probe convergence angle of \qty{31}{mrad}. The detector used was a sCMOS Ronchigram camera (Hamamatsu Orca v2). Using a probe defocus value of \qty{13}{nm}, 4D-STEM data were collected for every probe position in the scan, i.e.\ a diffraction pattern with reciprocal space co-ordinates ($k_x,k_y$) for every spatial position with real space co-ordinates ($x,y$). The beam current was \qty{30}{pA}, pixel dwell time \qty{5}{ms}, scan spacing \qty{0.047}{nm}, and scan size 128$\times$128 pixels (field of view of \qty{\sim6}{nm}). The time-averaged dose rate was thus \qty{5.2e6}{e/nm^2/s}. The region was scanned several times before the final acquisition, giving a total dose received by the area of \qty{\sim2.5e9}{e/nm^2}. 

The ptychographic reconstruction was performed in the open-source package \texttt{py4DSTEM}~\cite{savitzky2021py4dstem, varnavides2023iterative}. 
We used a single slice stochastic gradient descent algorithm with a batch size of 4096 probe positions.
The final complex object reconstruction used 100 iterations of step size 0.1 with probe position refinement.
A high-pass Butterworth filter was applied to the final object. 
The full reconstruction is shown in Fig.~S3.

The best-fit atomic positions were determined using custom Matlab code. 
First, we estimated the approximate atomic positions from local intensity maxima.
Next, we refined the positions by fitting a Gaussian distribution using nonlinear least squares.
Finally, we refined the positions by fitting a Gaussian distribution after subtracting the current estimated intensity from all neighboring peaks.
This last step was performed iteratively until convergence. 
In the reconstruction, the mean B--B distance corresponding to the bulk lattice was measured to be \qty{2.65\pm0.1}{\angstrom}. Figure~S3 shows the control region used for this measurement. Given the literature value of \qty{2.51}{\angstrom}~\cite{jain2013commentary}, the atomic positions in the reconstruction were scaled accordingly, from which all subsequent bond length measurements were then made. 

\begin{figure*}[]
  \centering
   \includegraphics[width=0.95\textwidth]{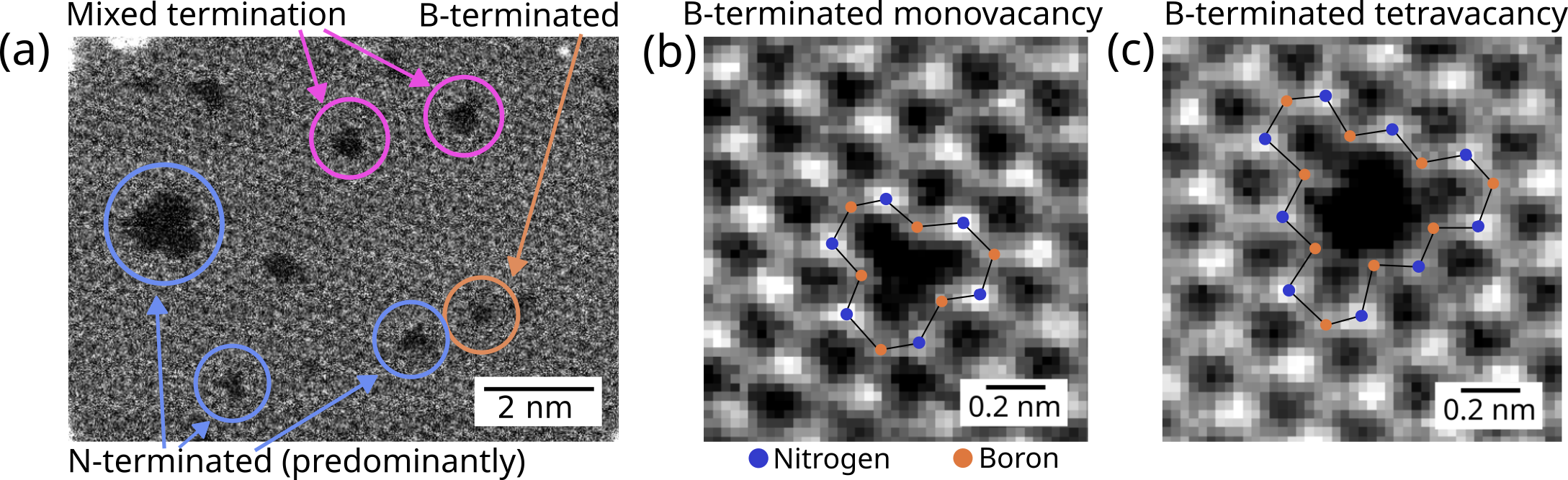}
    \caption{(a) HAADF-STEM survey showing vacancy defects in hBN after scanning at low dose rate. A number of defects are circled to indicate those with predominantly nitrogen termination (blue), boron termination (orange), and mixed edge termination (pink). (b-c) High magnification HAADF-STEM images of a monovacancy and a tetravacancy, respectively, acquired at high dose rate with edge atom models overlaid. These defects are boron terminated. The atom models shown here give approximate atom locations only.}\label{surveys}
\end{figure*}

\section{Results and discussion}\label{}

\subsection{Vacancy defects formed upon scanning at low dose rate vs.\ high dose rate}

Figure~\ref{surveys} shows HAADF-STEM images of hBN monolayer that had been pre-treated, as described in the methods, to produce a distribution of single-atom vacancies and larger vacancy defects for subsequent manipulation and analysis in STEM. In low dose rate scans, defects with different edge terminations were observed, as seen in Fig.~\ref{surveys}(a). Most of these defects were small (<24 atom vacancies) and include triangular defects with either nitrogen or boron atom termination, as well as hexagonal defects, which inherently feature alternating boron- and nitrogen-terminated edges. Edge termination was assigned based on Z-contrast, with nitrogen atoms appearing brighter than boron atoms. This contrast was used to identify the triangular nitrogen and boron atom sublattices, from which the edge termination of the geometric defects could be determined. Nitrogen- vs.\ boron-terminated triangular defects point in opposite directions.

Even though the imaging dose rates and total doses used for the survey images were relatively low ($\sim$\qty{e5}{e/nm^2/s} and $\sim$\qty{e7}{e/nm^2}, respectively), defect growth and movement between frames was observed. This illustrates the challenge of imaging defects in beam-sensitive materials non-invasively, while also demonstrating the capability to use the electron beam for intentional atom manipulation. In particular the larger defects are more unstable and thus expand more quickly, such as the larger defect highlighted on the left hand side of Fig.~\ref{surveys}(a). 
This defect is essentially a nitrogen-terminated triangle with truncated corners terminated with boron. Overall we infer a preference for nitrogen-terminated growth in the low-dose scans, although we note that defects with no obvious geometric preference also formed. 

In higher dose rate scans ($\sim$\qty{e7}{e/nm^2/s}), boron-ter\-minated defects were predominantly observed. Examples are given in the HAADF images of Figs.~\ref{surveys}(b) and (c), which show a boron-terminated monovacancy and tetravacancy, respectively. While the monovacancy may have already been present, the tetravacancy formed upon the high dose rate scanning. Out of 62 high dose rate scans in which defect edge termination could be clearly identified (each scan comprising multiple frames), 44 scans had at least one frame with a boron-terminated defect, while only 16 scans had at least one frame with a nitrogen-terminated defect. We note that identification of stable defect states was often challenging, since defects were typically very mobile during scanning, growing and/or changing shape from one frame to the next. However, in several scans taken at dose rates as high as \qty{e8}{e/nm^2/s}, even greater preference for boron vs.\ nitrogen termination was indicated. 

Our results thus reveal a preference for nitrogen atom ejection at the higher dose rates, resulting in the boron-terminated defects observed. Of these, the tetravacancies were the most stable during the high dose rate scans. 
Interestingly, in subsequent survey imaging at the lower dose rates, some of these boron-terminated tetravacancies appeared to have reconfigured, or may have been filled with carbon atoms~\cite{Park2021}. This could have occurred when the beam was blanked, or during the low dose rate scanning. Such reconfiguration supports the conclusion that different damage mechanisms and kinetics are at play depending on the imaging conditions. 

\begin{figure}
  \centering
   \includegraphics[width=0.48\textwidth]{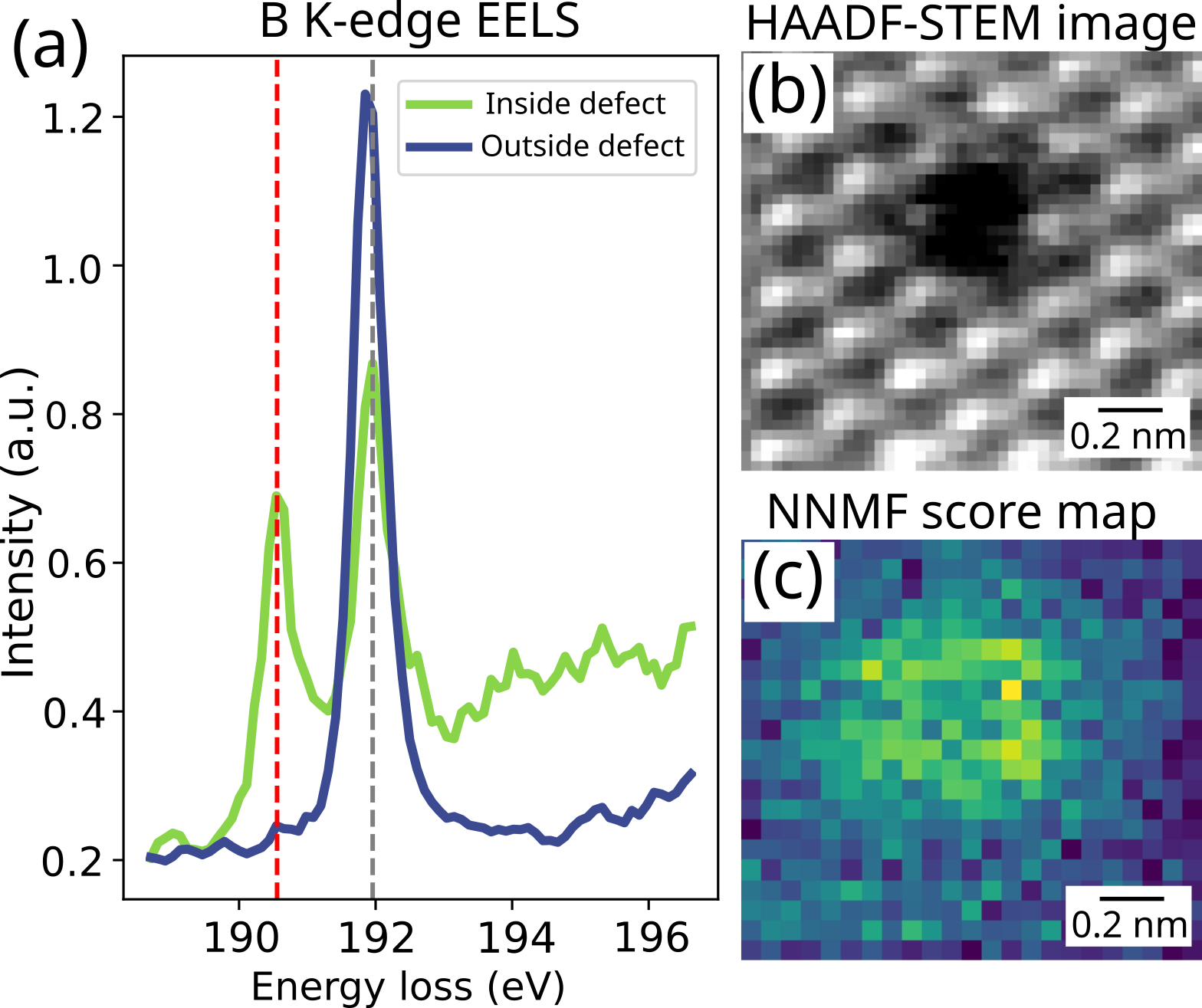}
    \caption{(a) Boron K-edge EEL spectrum imaging of a boron-terminated tetravacancy comparing the signal from inside the defect vs.\ outside the defect, normalized to the $\sigma$* peak (see Fig.~S2). The gray dashed line marks the energy of the $\pi$* peak from the bulk lattice, while the red dashed line indicates the energy at which a shifted pre-peak appears. (b) Corresponding HAADF-STEM image. (c) Score map for the pre-peak feature obtained by NNMF decomposition of the spectrum image. This map was used to create a mask to extract the individual spectra shown in (a).}\label{EELS}
\end{figure}

We additionally used STEM-EELS core-loss spectrum imaging to probe the defects, as shown in Fig.~\ref{EELS}. Since these were also high dose rate acquisitions (taken in parallel with the high-magnification HAADF-STEM images), boron-termination was prevalent. Figure~\ref{EELS}(a) shows boron K-edge EELS results for a boron-terminated tetravacancy, focusing on the $\pi$* region of the spectrum. The corresponding HAADF-STEM image is shown in Fig.~\ref{EELS}(b). A distinct pre-peak feature associated with the boron-terminated defect is found (marked with the dashed red line). This feature has been identified previously and analyzed in more detail by Cretu et al.~\cite{Cretu2015a}. Here, we use NNMF to identify the pre-peak in our dataset and to determine its spatial distribution. The resulting score map is shown in Fig.~\ref{EELS}(c), where the bright green pixels indicate the presence of the pre-peak. This map was then used to create a mask to generate the ``inside defect'' and ``outside defect'' spectra shown in Fig.~\ref{EELS}(a). More detail on the NNMF analysis can be found in Fig.~S1.

The pre-peak in the boron K-edge spectrum is essentially a chemical shift of the bulk $\pi$* peak due to the dangling bonds on the undercoordinated boron atoms, serving as a fingerprint for the boron-terminated tetravacancy~\cite{Cretu2015a}. Thus, both Z-contrast in HAADF-STEM and the pre-peak in the boron K-edge EELS can be used to determine the edge termination of the defect. The fact that the non-shifted (bulk) $\pi$* peak is detected inside the defect, as well as outside, is due to signal delocalization from the neighboring fully-coordinated boron atoms~\cite{Egerton2007}. 

Since near-edge structure is mainly affected by bonding with nearest neighbors, a distinct signature in the boron K-edge from a nitrogen-terminated tetravacancy is not expected (all the nitrogen atoms are fully coordinated)~\cite{Gao2019}. In the case of nitrogen-terminated tetravacancies, one can instead look for a pre-peak in the nitrogen K-edge~\cite{Gao2019,Suenaga2012}.


\begin{table*}
\begin{threeparttable}[width=2\linewidth,cols=7,pos=h]
    \caption{Overview of vacancy defect shapes and edge terminations induced by electron irradiation of monolayer hBN in TEM and STEM for various operation parameters, including the results of the present study.}\label{table}
    \begin{tabular*}{\tblwidth}{@{} LLLLLLL@{} }
        \toprule
  Defect shape & Edge termination & Mode & Voltage [kV] & Temperature [\qty{}{\degree C}] & Notes & References\\ 
        \midrule
        Triangle & N & (S)TEM$^1$ & 60/80/120 & RT & &  \citep{Jin2009,Alem2009,Ryu2015,Cretu2015a,Cretu2015} \\
        Triangle & N & STEM & 60 & RT & \textit{O$_2$ atmosphere} & \citep{Javed2024} \\
        Triangle & N & TEM & 60 & LN2 & & \citep{Cretu2015} \\
        Triangle & N & TEM & 30/60 & 400, 500 & &  \citep{Cretu2015} \\
        \midrule
        Triangle & B & STEM & 60 & RT & \textit{Higher dose rate, UHV} &  This work \\
        Triangle & B & STEM & 60/80 & 500 & &  \citep{Cretu2015a} \\
        \midrule
        Hexagon & Mixed & TEM & 80 & RT & \textit{Higher dose rate} & \cite{Gilbert2017} \\
        Hexagon & Mixed & TEM & 30/60/80 & 700--1000, 1200 & &  \citep{Cretu2015,Pham2016} \\
        \midrule
        Parallelogram & Mixed & TEM & 80 & 700--1000 & &  \citep{Pham2016} \\
        \midrule
        Circular & Mixed & STEM & 60 & RT & \textit{UHV} & \citep{Javed2024} \\
        Circular & Mixed & TEM & 60 & 800, 1000 & & \citep{Cretu2015} \\
        \bottomrule
    \end{tabular*}
     \begin{tablenotes}
            \item $^1$Z-contrast in STEM makes boron vs. nitrogen termination of the triangular defects straightforward to discern. This is not the case for TEM. Therefore, only those TEM results where the triangle edge terminations were confirmed by more detailed analysis are listed here.
        \end{tablenotes}
\end{threeparttable}
\end{table*}

\subsection{Contextual analysis of experimental parameters affecting defect growth energetics}
\label{lit-review}

As mentioned in the Introduction, various experimental parameters can influence the energetics of defect growth and hence steer the final defect shape and edge termination. Table~\ref{table} illustrates this point, comparing our observation of boron-atom termination under high dose rate STEM exposure with trends reported in prior studies. The table is organized according to defect shape: triangle, hexagon, parallelogram and circular. Results from both STEM and TEM experiments are given. In the STEM experiments, Z-contrast in HAADF means that boron and nitrogen can be easily differentiated, making edge terminations straightforward to assign. This is not the case in TEM mode, where differences in phase contrast between the two elements are much more subtle. Therefore, for the TEM experiments that report triangular defects (shapes that have edges terminated entirely by boron or nitrogen atoms), only those cases are listed in which further analysis was performed to determine the edge composition. The beam acceleration potentials used tend to be low (usually \qty{\leq 80}{kV}) to avoid excessive defect growth due to knock-on (elastic) collisions. 

Most studies have reported triangular defects that are nitrogen terminated, primarily in experiments performed at room temperature~\cite{Jin2009,Alem2009,Ryu2015,Cretu2015a,Cretu2015}, but also under cryogenic conditions and at elevated temperatures up to \qty{500}{\degree C}~\cite{Cretu2015}. The majority of these studies were conducted in TEM mode. However, in a particular set of TEM experiments performed at \qty{500}{\degree C}, a fraction of the triangular defects tended towards boron termination~\cite{Cretu2015a}. In the same work, the authors reported that under STEM imaging at this temperature, exclusively boron-terminated defects were formed. This marks a distinct shift from preferential boron atom removal (i.e. nitrogen termination) at lower temperatures to preferential nitrogen atom removal (boron termination) at the higher temperature.

Upon increasing the temperature even further, vacancy defects of other shapes can form, namely hexagons \cite{Cretu2015,Pham2016}, parallelograms~\cite{Pham2016}, and approximately circular shapes~\cite{Cretu2015}. Since these shapes exhibit alternating or mixed edge terminations, this suggests a regime where there is no longer a strong preference for the ejection of one element over the other. However, which of these non-triangular shapes is produced seems harder to predict, likely due to other varying parameters, such as the particular beam energy used, local charging of the hBN, residual species in the vacuum, and the electron dose rate. 

In fact, in recent STEM experiments comparing room temperature defect growth in UHV vs.\ oxygen atmosphere, it was found that circular pores formed in UHV, whereas upon introducing oxygen, nitrogen-terminated triangles were produced~\cite{Javed2024}. Initially this result appears to be at odds with our own observation of triangular defects forming in UHV STEM. However, the vacuum level reported in~\cite{Javed2024} is an order of magnitude lower than ours, which could help account for the difference. A change in the energetics of pore growth depending on vacuum quality and partial pressure of oxygen has also been observed for graphene~\cite{Leuthner2021}. And in another room temperature study on hBN, this time in TEM mode, it was found that non-triangular defects, in this case hexagons, could be grown by increasing the electron dose rate~\cite{Gilbert2017}. Therefore, even at room temperature, the energetics of pore growth can be strongly influenced by other experimental parameters. 

Placing the results of our STEM study in the context of the existing literature, we thus propose that increasing the dose rate had an equivalent effect on the defect growth process as increasing the temperature to \qty{500}{\degree C} did elsewhere \cite{Cretu2015a}. That is, both increasing the dose rate and increasing the temperature can drive a transition from preferential boron atom ejection to preferential nitrogen atom ejection, forming boron-terminated tetravacancies. Since we were operating at \qty{60}{kV}, i.e.\ below the elastic knock-on threshold for ejection of atoms from the pristine lattice, damage due to inelastic effects will dominate~\cite{Susi2019}. In the case of hBN, this can also include charging~\cite{Cretu2015}, which can be expected to have a strong dependence on electron dose rate. Beam-mediated processes that preferentially consume nitrogen, involving chemical reactions with residual species in the vacuum or on the sample surface, can also play a role~\cite{Cretu2015a}.

\subsection{Probing boron-terminated tetravacancies using electron ptychography}

\begin{figure*}
  \centering
   \includegraphics[width=0.95\textwidth]{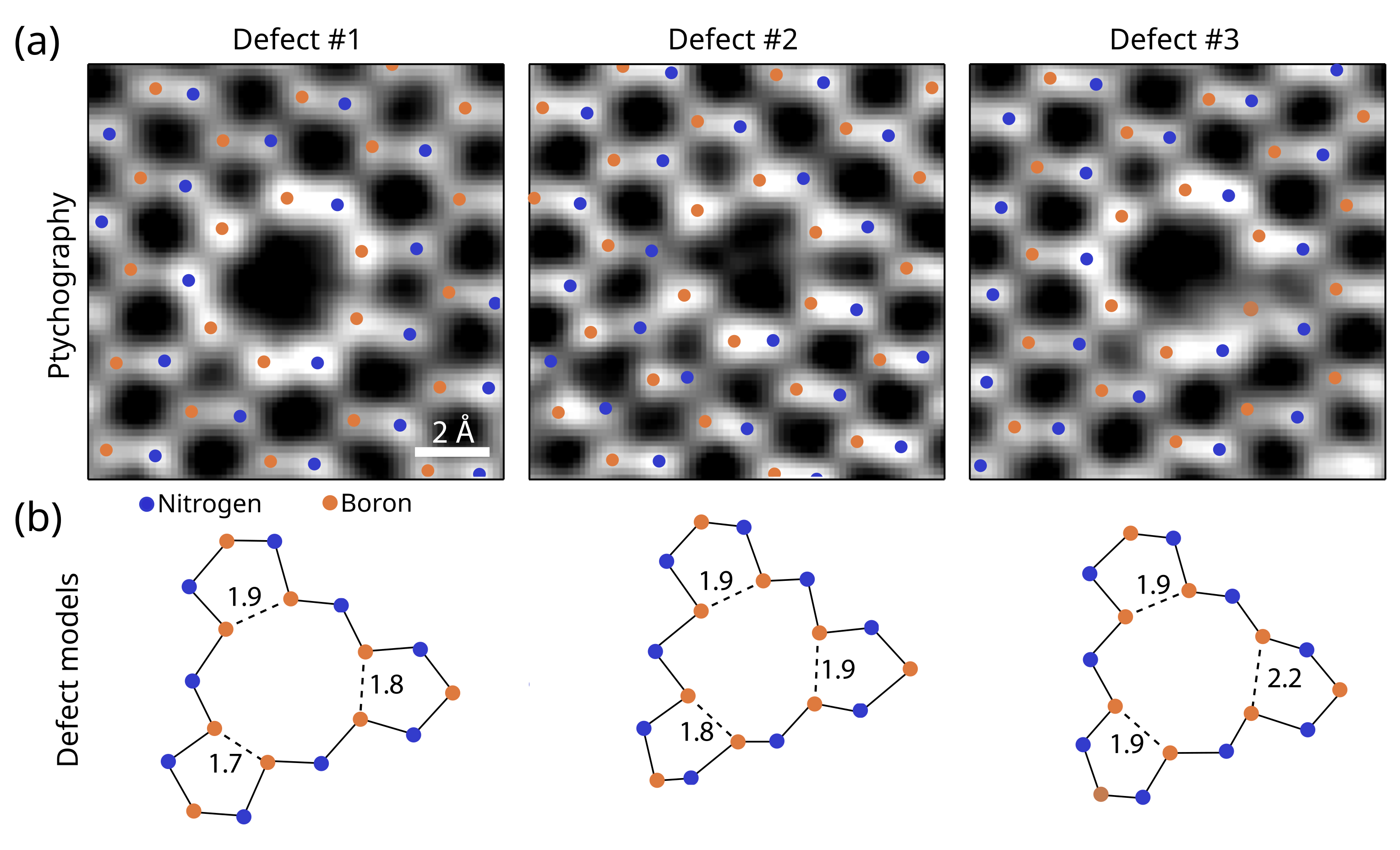}
    \caption{(a) Ptychographic reconstructions of boron-terminated tetravacancies in monolayer hBN with atomic map overlays. (b) Positions of the boron and nitrogen atoms around the defect edges as identified from the reconstructions. Dashed lines mark the B--B distances in the corner regions of the defects, measured in \text{\AA ngstr\"om} with a standard deviation of \qty{0.1}{\angstrom}.}
    \label{ptycho}
\end{figure*}

To further probe the structure and bonding of the boron-terminated tetravacancies, we performed electron ptychography. Fig.~\ref{ptycho}(a) summarizes the results, showing ptychographic reconstructions of three separate tetravacancies with atom map overlays. The atomic coordinates for the maps were determined using an iterative Gaussian distribution fitting method, as described in the methods. 

While one of the strengths of ptychography is its dose efficiency, we purposefully used higher doses for our acquisitions. As discussed previously, higher dose rates appear to drive the defect growth towards boron termination. Indeed, all of the tetravacancies observed in the ptychographic reconstructions are found to be boron terminated, consistent with our high dose rate observations using HAADF and EELS. The shape and edge composition of the tetravacancies determined by the ptychography is visualized more closely via the defect models shown in  Fig.~\ref{ptycho}(b).

Dashed lines in Fig.~\ref{ptycho}(b) mark the distances between the boron atoms in the corner regions of the triangular tetravacancies. The average B--B distance from these measurements is \qty{1.9(1)}{\angstrom}, which is significantly shorter than the B--B distance in the pristine lattice of \qty{2.5}{\angstrom}. Similar contraction has been revealed by HAADF-STEM imaging of boron-terminated tetravacancies elsewhere~\cite{Cretu2015} and has been attributed to direct bonding between the boron atoms in the corners upon structural relaxation of the defect~\cite{Cretu2015,Okada2009}. We note that lattice distortion is also observed more broadly in the reconstruction, as shown in Fig.~S3 (full field of view, encompassing all three defects analyzed in Fig.~\ref{ptycho}). The observed distortion may be an artifact caused by charging during the scan and could account for the more pronounced deformation of Defect \#2. 

An intriguing feature revealed by the ptychography is enhanced brightness around the defect perimeters. The enhancement is particularly marked for Defect \#1, where the nine atoms forming the ring-like structure (highlighted in the defect model) are those that appear brighter in the reconstruction. 
Since Z-contrast in the HAADF-STEM surveys did not show this effect, we rule out dopant species as the cause. 
In addition, since both the boron and nitrogen atoms exhibit the increased brightness, we conclude that the effect is not specific to a single atomic species. 
A possible explanation may be that we are observing the effect of atom movement due to instability at the open edges. However, it seems more likely that we are in fact mapping charge redistribution due to bonding, as recently investigated by electron ptychography of sulfur vacancies in monolayer molybdenum disulfide~\cite{Hofer2025}. The possibility to use electron ptychography to map bonding around defects in low-Z 2D materials such as hBN is a tantalizing prospect. 

\section{Conclusions}\label{}

We have investigated the effect of \qty{60}{kV} electron irradiation in STEM at low vs.\ high dose rate on the growth of vacancy defects in monolayer hBN. Over 60 datasets were analyzed to determine the dominant defect shapes and edge terminations. 
In low dose rate survey imaging, a mixture of triangular and hexagonal vacancy defect structures with varying edge termination were identified, whereas at higher dose rates, boron-terminated triangular defects were most commonly observed. Of the latter, boron-terminated tetravacancies were found to be the most stable. All experiments were performed in UHV with the samples held at room temperature. Contextualizing our findings within the existing literature, we conclude that preferential nitrogen atom ejection at the higher dose rates drives the formation of boron-terminated defects in a similar manner to increasing the sample temperature. In addition to Z-contrast in HAADF-STEM, we confirmed how near-edge structure in the boron K-edge can be used to assign boron termination. Finally, using 4D-STEM electron ptychography we detected a structural relaxation of the tetravacancies and revealed enhanced phase shift from the defect perimeters, likely hinting at charge redistribution at the defect edges to be studied in future work. 

\printcredits

\section*{Acknowledgments}

This work was funded in part by NSF Award No.\ 2110924. D.O.B. also acknowledges funding from the Department of Defense through the National Defense Science \& Engineering Graduate (NDSEG) Fellowship Program. 
S.M.R. and C.O. acknowledge support from the U.S. Department of Energy Early Career Research Award program.
SuperSTEM is the U.K.\ National Research Facility for Advanced Electron Microscopy, supported by the Engineering and Physical Sciences Research Council (EPSRC, UK) via grant numbers EP/W021080/1 and EP/V036432/1. 
The ion irradiation experiments were performed at the Biomolecular Nanotechnology Center, a core facility of the California Institute for Quantitative Biosciences.
Work at the Molecular Foundry was supported by the Office of Science, Office of Basic Energy Sciences, of the U.S. Department of Energy under Contract No.\ DE-AC02-05CH11231.
This research also used resources of the National Energy Research Scientific Computing Center (NERSC), a Department of Energy Office of Science User Facility using NERSC award BES-ERCAP0028035.
The authors thank Juan Carlos Idrobo and Georgios Varnavides for helpful discussions.

\end{document}


\begin{frontmatter}
    
\title{\texorpdfstring{Supplementary Information:\\ \vspace{2mm} Fabrication and characterization of boron-terminated tetravacancies in monolayer hBN using STEM, EELS and electron ptychography}}

\author[label1,label2,label3]{Dana O. Byrne}

\author[label3]{Stephanie M. Ribet}

\author[label4,label5]{Demie Kepaptsoglou}

\author[label4,label6]{Quentin M. Ramasse}

\author[label7]{Colin Ophus}

\author[label2,label3,label8]{\\Frances I. Allen}

\address[label1]{Department of Chemistry, University of California, Berkeley, CA 94720, USA}
\address[label2]{Department of Materials Science and Engineering, University of California, Berkeley, CA 94720, USA}
\address[label3]{Molecular Foundry, Lawrence Berkeley National Laboratory, Berkeley, CA 94720, USA}
\address[label4]{SuperSTEM Laboratory, SciTech Daresbury Campus, Daresbury, WA4 4AD, UK}
\address[label5]{School of Physics, Engineering and Technology \& JEOL Nanocentre, University of York, Heslington, YO10 5DD, UK}
\address[label6]{School of Chemical and Process Engineering \& School of Physics and Astronomy, University of Leeds, LS2 9JT, UK}
\address[label7]{Department of Materials Science and Engineering, Stanford University, Stanford, CA, 94305, USA.}
\address[label8]{California Institute for Quantitative Biosciences, University of California, Berkeley, CA 94720, USA}

\end{frontmatter}

\begin{figure*}[h]
 \centering
 \includegraphics[width=0.7\textwidth]{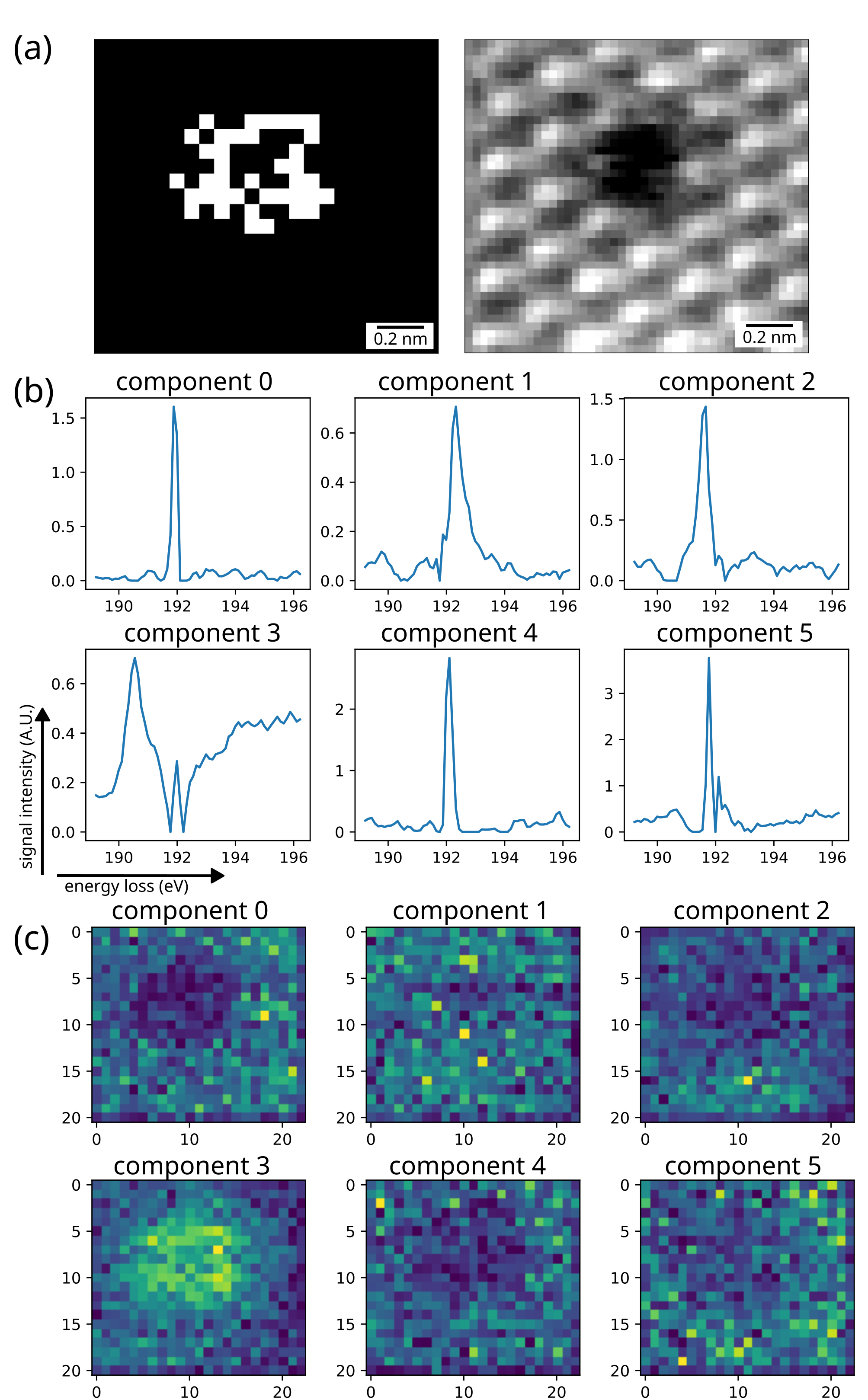}
 \caption{(a) Boolean mask used to extract the EEL spectra for the boron-terminated tetravacancy investigated in Fig.~2, and the corresponding HAADF. (b) Loadings (coefficients) obtained by NNMF decomposition of the spectrum-image dataset. (c) Corresponding score maps for these six components. Map 3 was used to generate the mask shown in (a).}
 \label{}
\end{figure*}

\begin{figure*}[h]
 \centering
 \includegraphics[width=0.95\textwidth]{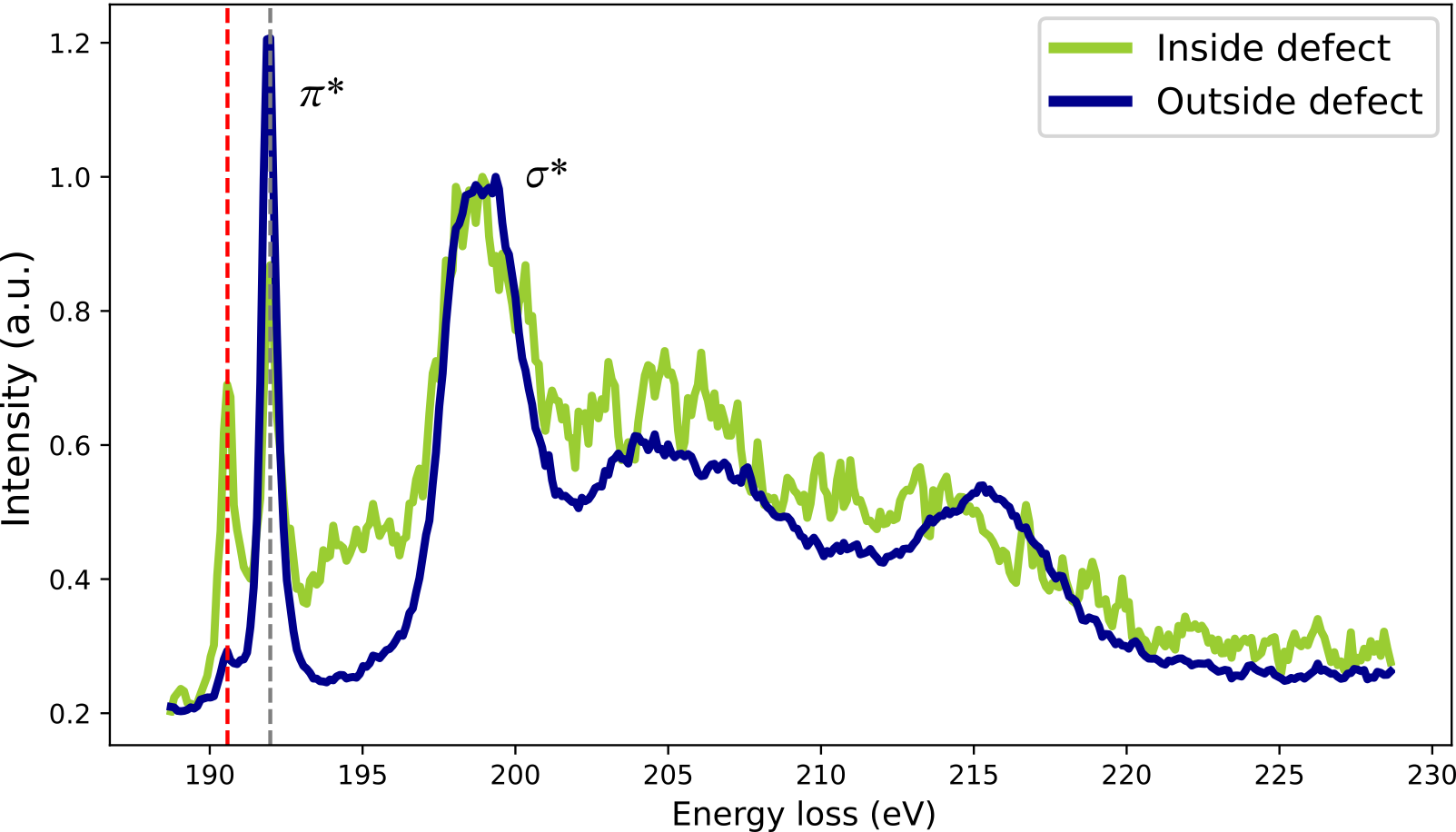}
 \caption{Full boron K-edge EELS of the boron-terminated tetravacancy shown in Fig.~2.}
 \label{}
\end{figure*}

\begin{figure*}[h]
 \centering
 \includegraphics[width=0.7\textwidth]{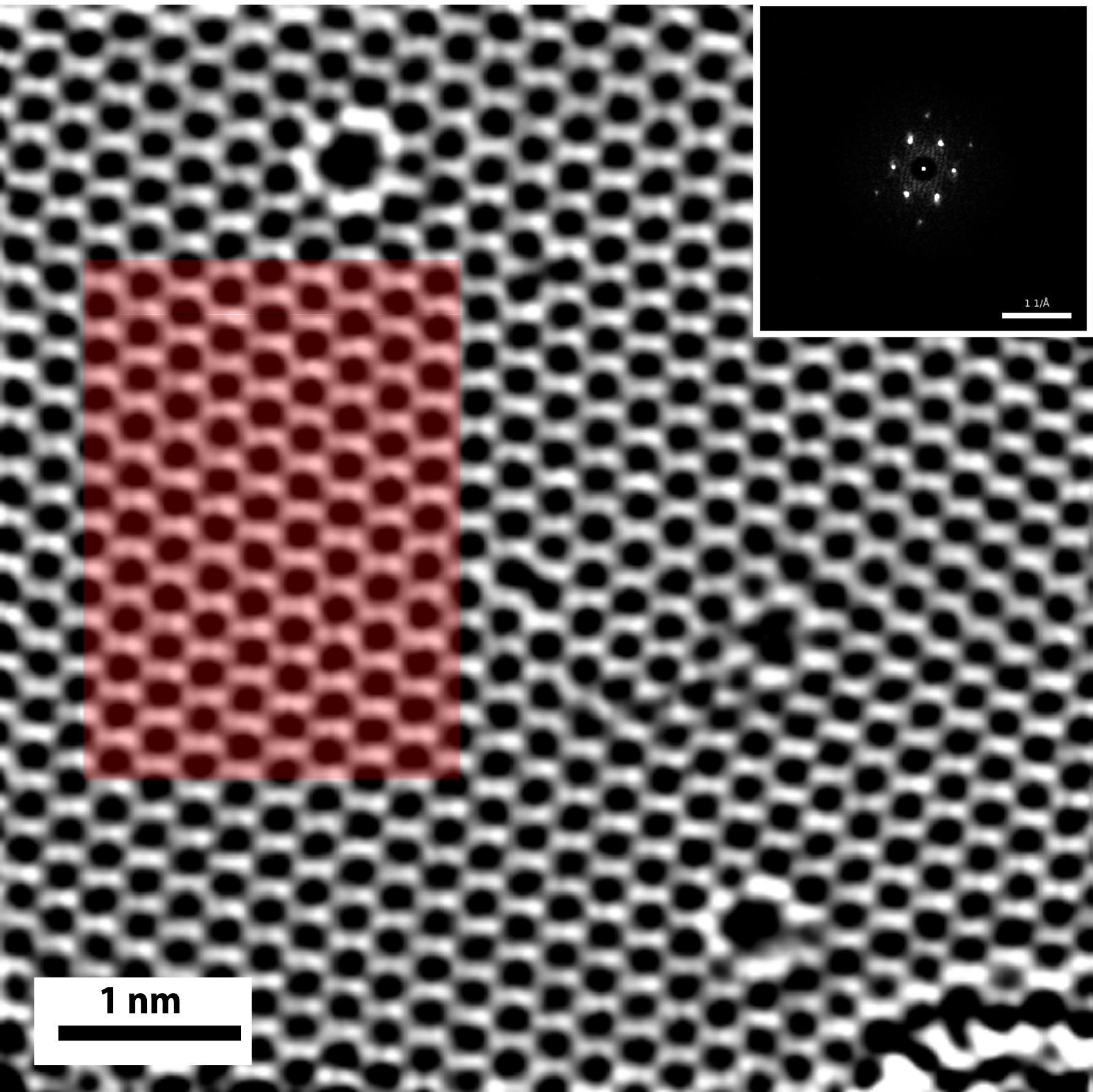}
 \caption{Full ptychography reconstruction of the monolayer hBN sample containing the defects analyzed in Fig.~3. The Fourier transform is provided in the inset. The red box highlights the control region used to measure the mean B--B distance in the bulk lattice. Distortion of the lattice is noticeable in some areas, which we attribute to charging.}
 \label{}
\end{figure*}